\documentclass[epj]{svjour}
\onecolumn
\usepackage{amssymb}

\begin{document} 

\title{Higher-order $\hbar$ corrections in the semiclassical quantization
 of chaotic billiards}

\author{Kirsten Weibert \and J\"org Main \and 
G\"unter Wunner}
\institute{Institut f\"ur Theoretische Physik 1, Universit\"at Stuttgart, 
D-70550 Stuttgart, Germany}

\date{\today}
%

\abstract{
In the periodic orbit quantization of physical systems, usually only the 
leading-order $\hbar$ contribution to the density of states is considered. 
Therefore, by construction, the eigenvalues following from semiclassical 
trace formulae generally agree with the exact quantum ones only to lowest 
order of $\hbar$.  In different theoretical work the trace formulae have
been extended to higher orders of $\hbar$.  The problem remains, however, 
how to actually calculate eigenvalues from the extended trace formulae 
since, even with  $\hbar$ corrections included, the periodic orbit sums 
still do not converge in the physical domain.  For {\em lowest-order} 
semiclassical trace formulae the convergence problem can be elegantly, and 
universally, circumvented by application of the technique of harmonic 
inversion.  In this paper we show how, for general scaling chaotic
systems, also {\em higher-order} $\hbar$ corrections to the Gutzwiller 
formula can be included in the harmonic inversion scheme, and demonstrate 
that corrected semiclassical eigenvalues can be calculated despite the 
convergence problem.  The method is applied to the open three-disk 
scattering system, as a prototype of a chaotic system.
} 

\maketitle

\section{Introduction}

The relation between the eigenvalue spectrum of a quantum system and the 
periodic orbits of the corresponding classical system is 
a question of fundamental importance for both integrable and chaotic
dynamical systems.
The well-established Gutzwiller trace formula \cite{Gut67,Gut71,Gut90}
for classically chaotic systems and its analogue for integrable systems,
the Berry-Tabor formula \cite{Ber76,Ber77}, provide the semiclassical
density of states in terms of a sum over all periodic orbits of the system.
However, each trace formula is only the leading-order term
of an expansion of the exact density
of states in powers of $\hbar$, and therefore in general
the resulting semiclassical eigenvalues
are only approximations to the exact quantum ones.
In recent years, two basic methods have been developed for determining 
higher-order $\hbar$ corrections to the Gutzwiller trace formula in terms 
of periodic orbit contributions \cite{Alo93,Gas93,Vat94,Vat96,Ros94,Gre01}.
Unfortunately, even with $\hbar$ corrections included, the trace formulae
usually suffer from being divergent in the region where the physical
eigenvalues or resonances are located.
For specific systems, higher-order $\hbar$ corrections to the
semiclassical eigenvalues have explicitly
been calculated by cycle expansion techniques \cite{Alo93,Vat96,Ros94}.
However, this method is applicable only  to systems with special features, 
namely to hyperbolic systems with a known complete symbolic dynamics.

Recently, it has been demonstrated how the convergence problems
of the semiclassical trace formulae can be circumvented by the application
of harmonic inversion techniques \cite{Mai97b,Mai98,Mai99a}.
The harmonic inversion method is capable of extracting semiclassical
eigenvalues from a finite set of periodic orbits with very high
precision and resolution.
In contrast to other semiclassical methods, harmonic inversion 
does not require any special properties of the system, and can therefore
be applied to a wide range of physical systems.
In Refs.~\cite{Mai99a,Mai98c,Wei00} a general procedure has been
developed for  including higher-order $\hbar$ corrections to the
trace formulae in the harmonic inversion scheme.
So far, this method has only been tested for an integrable system, viz.\
the circle billiard.
The general procedure, however, does not depend on the type of the 
underlying classical dynamics,
and is  applicable also to chaotic systems.
In this paper, we demonstrate how the method works for chaotic systems,
and  apply it to the open three-disk scatterer, which has become a
standard example for the semiclassical quantization of chaotic systems
\cite{Mai98,Mai99a,Gas89,Cvi89,Eck93,Eck95,Wir99}.

The harmonic inversion method is used in two directions:
First, we carry out a harmonic analysis of the spectrum of the differences 
between the exact (complex) quantum  eigenvalues
and the semiclassical resonances of the three-disk system. We show how 
this enables one to determine, for each orbit,
the first-order $\hbar$
correction term (and, in principle, all higher-order correction terms) 
to the Gutzwiller formula. We confirm our results by comparing
with the  values calculated by
a specialization of an analytical approach developed by
Vattay and Rosenqvist \cite{Vat94,Vat96}
for two-dimensional billiards \cite{Ros94}.
Second, we take the analytical correction terms to the Gutzwiller formula 
and  compute, using the classical periodic orbit data and harmonic inversion,
the first-order $\hbar$ corrections to the semiclassical resonances of the 
three-disk system.
Thus we illustrate that corrected semiclassical eigenvalues
can be calculated by harmonic inversion despite the convergence problems
of the trace formulae.
We compare the zeroth and first-order approximations to the resonances
with the exact quantum eigenvalues, and can quantitatively assess 
the increase in accuracy  produced by including the next-order corrections.

\section{Higher-order $\hbar$ corrections to Gutzwiller's trace formula}
\label{hbarsec1}
Gutzwiller's trace formula for chaotic systems gives a semiclassical 
approximation to the response function (i.e., the trace of the Green's 
function) of a quantum system in terms of the periodic orbits of the 
corresponding classical system. 
The semiclassical response function
consists of a smooth background, and an oscillating part
$g(E)=\bar g(E)+g^{\rm osc}(E)$, where the oscillating part is given by
\cite{Gut90,Bal74}
\begin{equation}
\label{Gutz}
 g^{\rm osc}(E) = -{\rm i}\sum_{\rm po}
 {T_{\rm po}\over r|\det (M_{\rm po}-{\bf 1})|^{1/2}}
 \exp\left[{\rm i}\left({S_{\rm po}\over\hbar}
   - \mu_{\rm po}{\pi\over 2}\right)\right] \; .
\end{equation}
The sum in (\ref{Gutz}) runs over all periodic orbits (po) of the system,
including multiple traversals.
Here, $T_{\rm po}$ and $S_{\rm po}$ are the period and the action of the orbit,
$M_{\rm po}$ and $\mu_{\rm po}$ denote the monodromy matrix and the Maslov index, 
and the repetition number $r$ counts the traversals of the 
underlying primitive orbit
(``primitive'' means that there is no sub-period).
The semiclassical density of states $\rho(E)$ is related to the
response function via
\begin{equation}
 \rho(E) = -{1\over\pi}\, {\rm Im}\, g(E)\; .
\end{equation}
In general, the semiclassical eigenvalues or resonances obtained from the
Gutzwiller formula agree with the exact quantum ones only in 
leading order of $\hbar$.
To improve the accuracy of the semiclassical eigenvalues, higher-order
$\hbar$ correction terms to the Gutzwiller formula have to be included. 
Two different methods for the calculation of 
such higher-order $\hbar$ terms have been derived for chaotic systems, one
by Gaspard and Alonso \cite{Alo93,Gas93}, 
and the other by Vattay and Rosenqvist \cite{Vat94,Vat96,Ros94}.
The latter method has been specialized to two-dimensional chaotic billiards 
in Ref.~\cite{Ros94}.
An extension of the method of Gaspard and Alonso has recently been published
in Ref.~\cite{Gre01}.
We will adopt the method of Vattay and Rosenqvist to compute 
the first-order $\hbar$ corrections to the semiclassical resonances
of the open three-disk scatterer.

Vattay and Rosenqvist give a quantum generalization of the Gutzwiller
formula, which is of the form 
\begin{equation}
 g(E)=\bar g(E)+{1\over {\rm i}\hbar} 
 \sum_p \sum_l \left(T_p(E)-{\rm i}\hbar
  {{\rm d}\ln R^l_p(E)\over {\rm d}E}\right)
 \sum_{r=1}^\infty \left(R^l_p(E)\right)^r
 \exp{\left({{\rm i}\over\hbar} rS_p(E)\right)}.
\label{qmGutzw1}
\end{equation}
The first sum runs over all {\it primitive} periodic orbits; 
$T_p$ and $S_p$ are the 
traversal time and the action of the periodic orbit, respectively.
The sum over $r$ corresponds to multiple traversals of the primitive
orbit.
The quantities $R^l_p$ are associated with the local
eigenspectra determined by the local 
Schr\"odinger equation in the neighbourhood of the periodic orbits.
An expansion of the quantities $R^l_p$ in powers of $\hbar$,
\begin{eqnarray}
R^l(E)&=&\exp\left\{\sum_{m=0}^\infty \left({{\rm i}\hbar\over2}\right)^m
 C_l^{(m)}\right\} \nonumber \\
&\approx& \exp\left(C_l^{(0)}\right)\left(1+{{\rm i}\hbar\over 2}C_l^{(1)}
+\dots \right)\;,
\label{defCl1}
\end{eqnarray}
yields the $\hbar$ expansion of the generalized trace formula 
(\ref{qmGutzw1}). 
For two-dimensional hyperbolic systems, the zeroth-order terms are given by
\begin{equation}\label{cl0}
   \exp\left(C_l^{(0)}\right)
 = {{\rm e}^{{\rm i}\mu_p\pi/2}\over |\lambda_p|^{1/2}\lambda_p^l}\;,
\end{equation} 
where $\mu_p$ and $\lambda_p$ are the Maslov index and the expanding
stability eigenvalue (i.e., the stability eigenvalue with an absolute 
value larger than one) of the orbit, respectively.
By summation over $l$, the Gutzwiller trace formula is regained as
zeroth-order approximation to Eq.~(\ref{qmGutzw1}).
If the zeroth-order terms do not depend on the energy, as is the
case for billiard systems, the first-order correction to the Gutzwiller 
formula can be written as
\begin{equation}\label{g1_explizit2}
 g_1(E)={1\over {\rm i}\hbar} 
 \sum_{\rm po} \sum_l {T_{\rm po}(E)\over r}
 \exp\left(C_l^{(0)}\right)\ {{\rm i}\hbar\over 2}\;C_l^{(1)}
 \exp\left({{{\rm i}\over\hbar} S_{\rm po}(E)}\right)\; ,
\end{equation}
where the first sum in Eq.~(\ref{g1_explizit2}) 
now runs over {\it all} periodic orbits, including multiple traversals,
and $r$ is the repetition number with respect to the underlying primitive
orbit.

An explicit recipe for the calculation of the correction terms $C_l^{(1)}$ 
for two-di\-mens\-ion\-al chaotic billiards was given in Ref.~\cite{Ros94}.
The correction terms must in general be calculated numerically
from the periodic orbit data.
A numerical code which determines
the first-order corrections for two-dimensional chaotic billiards 
can also be found in Ref.~\cite{Ros94}.
We have used that code to compute the correction terms $C^{(1)}_l$ for
the three-disk system for a comparison with the correction terms 
calculated by harmonic inversion.

\section{The open three-disk scatterer}
As a model system for the calculation of higher-order $\hbar$
corrections to the Gutzwiller formula by harmonic inversion,
we consider the open three-disk system, which consists of three equally
spaced hard disks of unit radius.
This system, in particular the case of the relatively large disk separation 
$d=6$, has served as an archetype  for the application of semiclassical 
quantization techniques in many investigations in  recent years
\cite{Mai98,Mai99a,Cvi89,Eck93,Eck95,Wir99}.
We will consider the case $d=6$, as well as the small separation
$d=2.5$.
In our calculations, we make use of the symmetry reduction 
of the three-disk system introduced
in Refs.~\cite{Cvi89,Cvi93} and concentrate on states of the $A_1$ subspace.

As for all billiard systems, the shape of the periodic orbits in the 
three-disk system is independent of the wave number $k=\sqrt{2mE}/\hbar$, 
and the action scales as
\begin{equation}
 S/\hbar= k s,
\end{equation}
where the scaled action $s$ is equal to the physical length of the orbit.
We consider the density of states as a function of the wave number
\begin{equation}
 \rho(k) = -{1\over\pi}\, {\rm Im}\ g(k) \; ,
\end{equation}
with a scaled response function $g(k)$.
Since the wave number $k$ is proportional to $\hbar^{-1}$, it can be 
considered as an effective Planck constant,
\begin{equation}
\label{hbar_eff}
 k = \hbar_{\rm eff}^{-1} \; .
\end{equation}
The $\hbar$ expansion of the exact quantum response function can therefore 
be written as a power series in $k^{-1}$:
\begin{equation}
 g(k) = \bar g(k)+g^{\rm osc}(k)
\end{equation}
with
\begin{equation}
   g^{\rm osc}(k)
 = \sum_{n=0}^\infty g_n(k)
 = \sum_{n=0}^\infty {1\over k^{n}} \sum_{\rm po} 
   {\cal A}_{\rm po}^{(n)} {\rm e}^{{\rm i}s_{\rm po}k} .
\label{g_hbar_series}
\end{equation}
The second sum runs over all periodic orbits including multiple traversals. 
The zeroth-order amplitudes ${\cal A}_{\rm po}^{(0)}$ correspond to the
Gutzwiller formula, whereas for $n>0$, the amplitudes 
${\cal A}_{\rm po}^{(n)}$ give the $n^{\rm th}$-order corrections 
$g_n(k)$ to the response function.

Applying the Gutzwiller trace formula to the (symmetry reduced) three-disk
system yields for the zeroth-order amplitudes in Eq.~(\ref{g_hbar_series}) 
($A_1$ subspace)
\begin{eqnarray}
\label{scGutz}
{\mathcal A}_{\rm po}^{(0)}  &=& 
 -{\rm i}\sum_{\rm po} 
    {s_{\rm po}\, {\rm e}^{-{\rm i}{\pi\over 2}\mu_{\rm po}}
    \over r |\det(M_{\rm po}-1)|^{1/2}} \nonumber \\
&=& -{\rm i} \sum_{\rm po} (-1)^{l_s} {s_{\rm po}\over r 
    |(\lambda_{\rm po} -1)({1\over \lambda_{\rm po}}-1)|^{1/2}} \; ,
\label{gw3disk}
\end{eqnarray}
where $M_{\rm po}$ is the monodromy matrix of the orbit,
$l_s$ is the symbol length,  $s_{\rm po}$  the scaled action,
 and $\lambda_{\rm po}$  the expanding 
stability eigenvalue of the orbit. The Maslov index $\mu_{\rm po}$ for
this system  is  given by $2 l_s$.
The quantity $r$ designates the
repetition number with respect to the corresponding primitive orbit.
The first-order amplitudes of the $\hbar$ expansion (\ref{g_hbar_series})
following from Eq.~(\ref{g1_explizit2}) read
\begin{equation}\label{A1_explizit}
{\mathcal A}_{\rm po}^{(1)}=
{s_{\rm po}\over r}
\sum_l{(-1)^{l_s}\over |\lambda_{\rm po}|^{1/2}\lambda_{\rm po}^l}\,
{C_l^{(1)}\over 2\hbar k}\; . 
\end{equation}
Since the terms $C_l^{(1)}$ are proportional to the momentum $\hbar k$,
as was shown in Ref.~\cite{Ros94},
the amplitudes are independent of the scaling parameter $k$.
The correction terms $C_l^{(1)}$ have to be determined numerically.
We use the code developed by Rosenqvist and Vattay \cite{Ros94,Vat}. 
The code requires the flight times between the bounces and the reflection 
angles as an input.  These parameters have to be calculated
numerically for each periodic orbit.
As the contributions to the amplitude (\ref{A1_explizit}) for
different $l$ are proportional to $|\lambda_{\rm po}|^{-l-{1\over 2}}$, the sum
over $l$ converges fast if the absolute value of the stability eigenvalue
$\lambda_{\rm po}$ is large. 
For most orbits, the leading term $l=0$ turns out to be already sufficient. 
It is only for the very shortest orbits that terms of higher order in $l$ 
have to be included to ensure convergence of the sum to within, say, 3 
significant digits.

\section{Harmonic analysis of the quantum spectrum}
\label{hbar2}
\subsection{Theory}
In Refs.~\cite{Mai99a,Wei00} it was demonstrated that the amplitudes
${\cal A}_{\rm po}^{(n)}$ of the $\hbar$ 
expansion (\ref{g_hbar_series}) can be obtained by a harmonic inversion
analysis of the exact quantum
spectrum.
The general procedures do not depend
on any special properties of the system and can be 
applied to both integrable and chaotic systems.
In Ref.~\cite{Wei00}, they were tested for
the circle billiard, as an example of an integrable system.
We will now use the same procedures for the open
three-disk system, as a representative of a chaotic system.

We start by briefly recapitulating the main ideas of the procedures developed
in Refs.~\cite{Mai99a,Wei00}.
The exact quantum mechanical response function, in terms of the wave number 
$k$, can be written as
\begin{equation}
\label{gqm}
 g^{\rm qm}(k) = \sum_j {m_j\over k-k_j+{\rm i}0} \; ,
\end{equation}
where the $k_j$ are the exact eigenvalues or resonances of $k$, and $m_j$ are
their multiplicities.
Eq.~(\ref{g_hbar_series}) gives the $\hbar$ expansion of (\ref{gqm})
in terms of periodic orbit contributions.
The first-order amplitudes ${\mathcal A}_{\rm po}^{(0)}$ of the expansion 
(\ref{g_hbar_series}) can be determined by adjusting the exact response 
function  (\ref{gqm}) to the form of the semiclassical approximation
\begin{equation}\label{g_allg_semi}
 g^{\rm osc}(k) \approx \sum_{\rm po} {\mathcal A}_{\rm po}^{(0)}
 {\rm e}^{{\rm i}ks_{\rm po}}
\end{equation}
by harmonic inversion \cite{Mai97a}.
Note that for chaotic billiards the leading-order
amplitudes ${\cal A}_{\rm po}^{(0)}$ 
as well as the higher-order amplitudes ${\cal A}_{\rm po}^{(n)}$ 
are independent of the wave number $k$.
In a direct harmonic analysis of the quantum spectrum 
only the zeroth-order term
of the expansion (\ref{g_hbar_series}) 
fulfills the ansatz for the harmonic inversion procedure.
The higher-order terms act as a kind of weak noise which is separated
from the zeroth-order ``signal'' by the harmonic inversion procedure.
The direct harmonic analysis of the quantum signal 
will therefore yield exactly the lowest-order amplitudes
${\cal A}_{\rm po}^{(0)}$.

The $n^{\rm th}$-order amplitudes ${\cal A}_{\rm po}^{(n)}$ can be determined
if the exact eigenvalues $k_j$ and their $(n-1)^{\rm st}$-order approximations 
$k_{j,n-1}$ are given.
The $(n-1)^{\rm st}$-order approximation to the response function
can be written in the form  (\ref{gqm}), with $k_j$ replaced with
the approximation $k_{j,n-1}$.
One can then calculate the difference between the exact quantum mechanical
and the $(n-1)^{\rm st}$-order response function and compare it with 
the expression resulting from the expansion (\ref{g_hbar_series}),
\begin{equation}
\label{Dg_n}
   g^{\rm qm}(k) - \sum_{m=0}^{n-1} g_m(k)
 = \sum_{m=n}^\infty g_m(k)
 = \sum_{m=n}^\infty {1\over k^{m}} \sum_{\rm po} 
   {\cal A}_{\rm po}^{(m)} {\rm e}^{{\rm i}s_{\rm po}k} .
\end{equation}
The leading-order terms in (\ref{Dg_n}) are $\sim k^{-n}$, i.e.,
multiplication by $k^n$ yields
\begin{equation}
\label{G_n}
   k^n\left[g^{\rm qm}(k) - \sum_{m=0}^{n-1} g_m(k)\right]
 = \sum_{\rm po} {\cal A}_{\rm po}^{(n)} {\rm e}^{{\rm i}s_{\rm po}k}
   + {\cal O}\left({1\over k}\right)\; .
\end{equation}
The right-hand side of (\ref{G_n}) now has assumed a form which is again suited
to the harmonic inversion procedure. More precisely,
the harmonic inversion of the weighted difference spectrum (\ref{G_n})
will yield the 
periods $s_{\rm po}$ and the $n^{\rm th}$-order amplitudes 
${\cal A}_{\rm po}^{(n)}$ of the $\hbar$ expansion (\ref{g_hbar_series}).

\subsection{Application to the three-disk scattering system}

\begin{figure}
\vspace{9.5cm}
\includegraphics{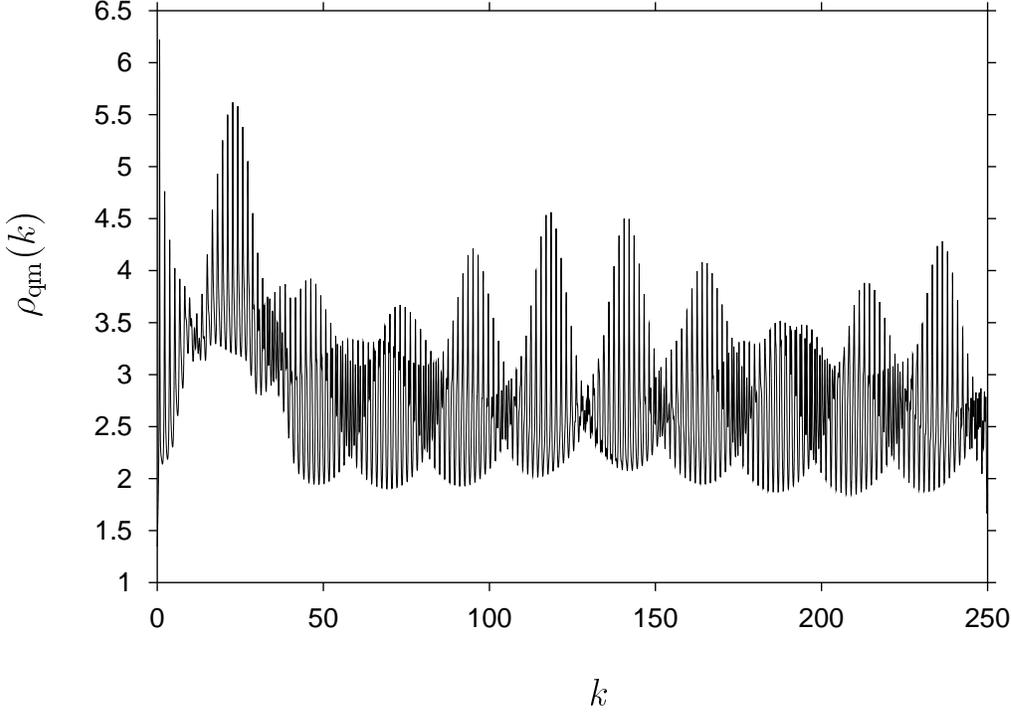}
\caption{Quantum mechanical density of states 
$\rho(k)=(-1/\pi)\,{\rm Im}\,g(k)$
of the three-disk system with disk separation $d=6$ ($A_1$ subspace) 
as a function of real values of the wave number $k$. Only resonances
of the four leading bands have been included. [Data courtesy of A. Wirzba.]}
\label{rho.fig}
\end{figure}

We now apply the procedure to
the three-disk system with disk separation $d=6$.
As a  first step we  perform a harmonic analysis of the exact quantum spectrum 
to obtain the leading-order ($n = 0$) periodic orbit contributions to the
density of states. The exact quantum values for $d=6$ were taken from Wirzba 
\cite{Wir99,Cvi97,Wir}.
Figure \ref{rho.fig} shows the quantum mechanical density of states
$\rho(k)=(-1/\pi)\,{\rm Im}\,g(k)$ 
for real values of the wave number $k$
resulting from the four leading bands of the 
$A_1$ subspace. Note that this set of resonances is of course not complete
as the subleading bands with large negative imaginary part are not included.

\begin{figure}
\vspace{13.2cm}
\includegraphics{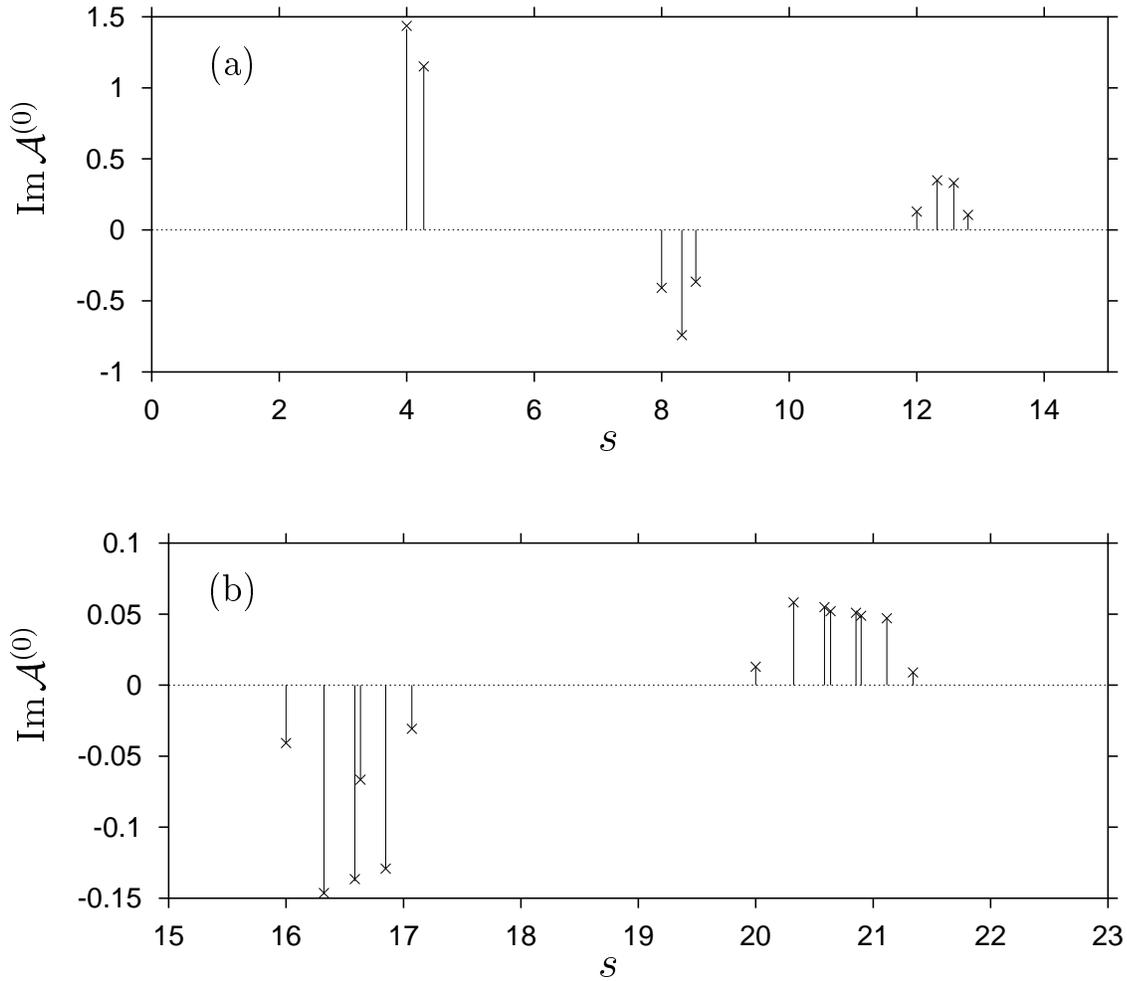}
\caption{Imaginary parts of the amplitudes of the 
leading-order ($n=0$)
periodic orbit contributions to the density of states of the three-disk 
system with disk separation $d=6$ as a function of the scaled actions
of the symmetry reduced orbits.
Solid lines: semiclassical amplitudes ${\mathcal A}_{\rm po}^{(0)}$
versus scaled actions of the symmetry reduced orbits, calculated directly from 
classical mechanics. Crosses: results from the harmonic inversion of the exact 
quantum spectrum ($A_1$ subspace). }
\label{0.Ordnung}
\end{figure}

The spectrum in Figure \ref{rho.fig} served as    the signal for the harmonic 
inversion procedure.
The results of the analysis
turned out  to be more accurate if the lowest part of the signal, determined
by the ``most quantum'' resonances with very small real part, is cut off.
We analyzed the spectrum in the range 
${\rm Re}\,k\in [50,250]$ to obtain the periodic orbit contributions
in two different intervals of the scaled action.
The results are presented in Figure \ref{0.Ordnung}.
The solid lines give the sizes of  
the imaginary parts of the semiclassical amplitudes
${\mathcal A}_{\rm po}^{(0)}$ 
calculated directly from Eq.~(\ref{scGutz}) using the
classical periodic orbit data, 
as a function of the scaled action of the orbits. 
The crosses show the amplitudes resulting from the
 harmonic inversion of the quantum spectrum.
Note the different scales of the two plots. 
The  results of the harmonic inversion
 are seen to be in excellent agreement with
those from the classical calculations of the amplitudes
${\mathcal A}_{\rm po}^{(0)}$   entering into
Gutzwiller's trace formula, clearly confirming the validity of the latter.

\begin{figure}
\vspace{9.5cm}
\includegraphics{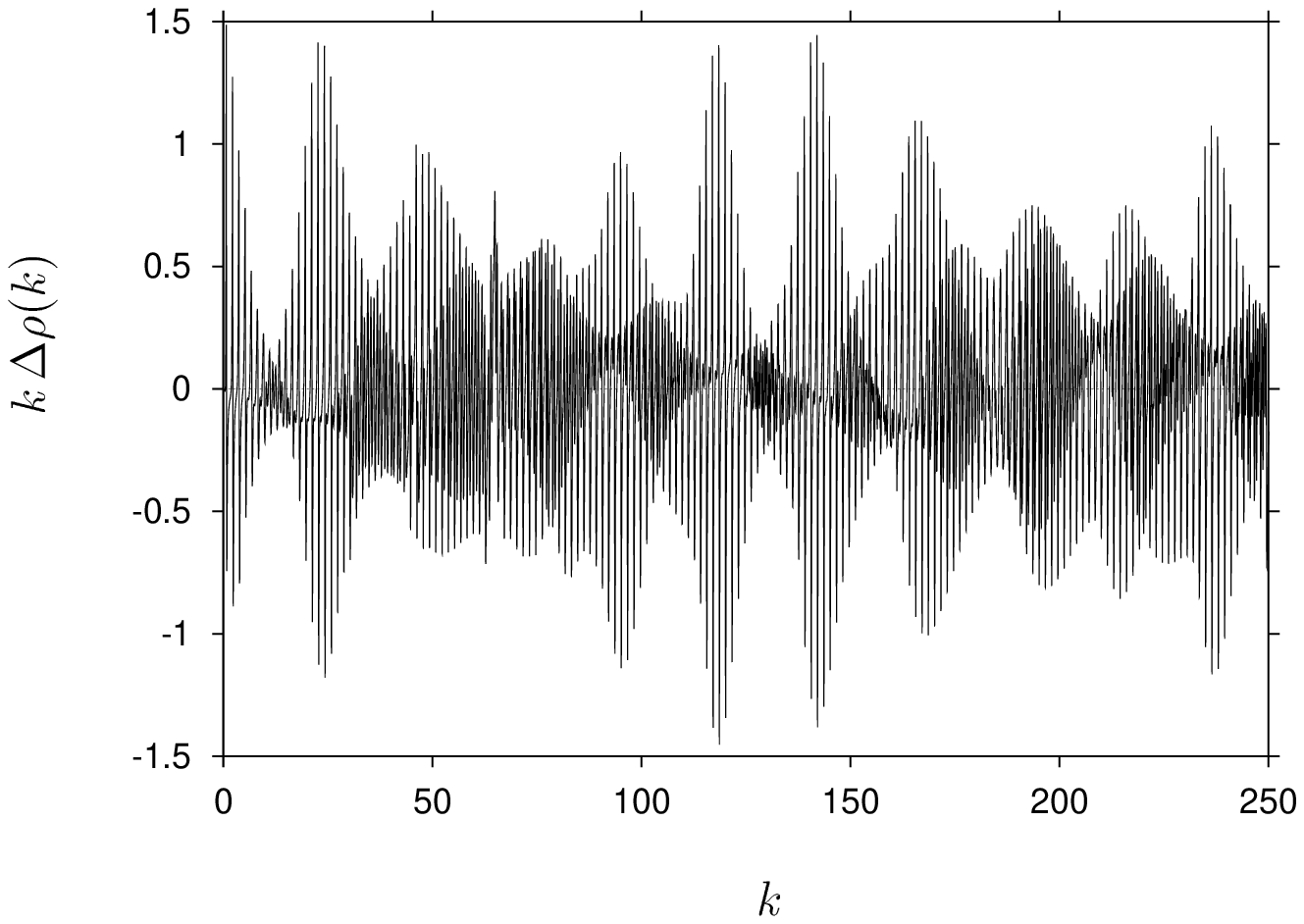}
\caption{Three-disk system with disk separation $d=6$: Weighted
difference spectrum $k\, \Delta\rho(k)=k(\rho_{\rm qm}(k)-\rho_{\rm sc}(k))$ 
between the quantum mechanical and the semiclassical density of states 
($A_1$ subspace) as a function of the wave number $k$.}
\label{delta_rho.fig}  
\end{figure}

In a second step, we now determine the next-to-leading-order $\hbar$ 
corrections to the Gutzwiller trace formula for the three-disk system 
by the harmonic analysis of the difference spectrum between the exact 
quantum resonances and the semiclassical resonances of the $A_1$ subspace.
The semiclassical resonances for disk separation $d=6$
had been calculated by Wirzba 
\cite{Wir99,Cvi97,Wir} from a  cycle expansion of the  
Gutzwiller-Voros zeta function.
[Since the Gutzwiller-Voros zeta function is directly related to
the Gutzwiller trace formula without further approximations, 
the semiclassical resonances resulting from both expressions 
will  be the same.]
The weighted difference spectrum is shown in Figure \ref{delta_rho.fig}.
Note that due to the limited radius of convergence of the cycle expansion 
only resonances with ${\rm Im}\,k\gtrsim -0.8$ were available and
could be included in the signal.

\begin{figure}
\vspace{13.2cm}
\includegraphics{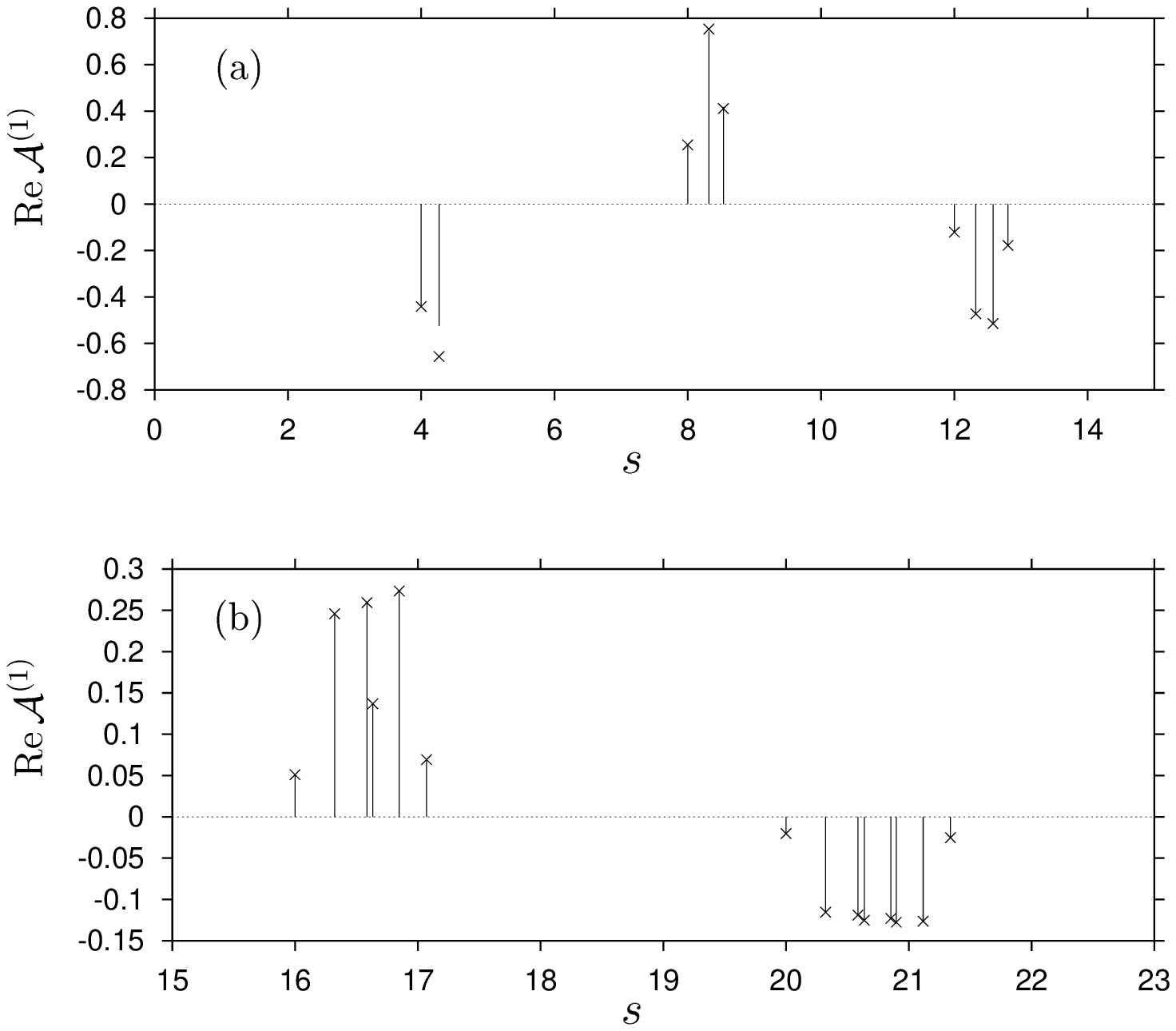}
\caption{
First-order $\hbar$ corrections to the trace formula of the 
three-disk system with disk separation $d=6$ as a function of the
scaled actions of the symmetry reduced orbits. Solid lines: first-order 
amplitudes following from a direct evaluation of Eq.~(\ref{A1_explizit}). 
Crosses: results from the harmonic inversion 
of the difference spectrum between exact quantum resonances and 
semiclassical cycle expansion values ($A_1$ subspace).
}
\label{1.Ordnung}
\end{figure}

In Figure \ref{1.Ordnung}, the crosses designate the results 
for the first-order amplitudes ${\cal A}_{\rm po}^{(1)}$ obtained
from the harmonic inversion of the difference spectrum shown in
Fig.~\ref{delta_rho.fig}, which was analyzed in the region 
${\rm Re}\,k\in [100,250]$.
For comparison, we also determined the first-order amplitudes
${\cal A}_{\rm po}^{(1)}$ for each orbit following the method of Vattay 
and Rosenqvist described above (see Eq.~(\ref{A1_explizit})).
These results are represented in Figure \ref{1.Ordnung} by solid lines.

For almost all orbits, the harmonic inversion results for 
${\cal A}_{\rm po}^{(1)}$ are seen to be in perfect agreement with the 
amplitudes calculated by the method of Refs.~\cite{Vat94,Vat96,Ros94}.
There is, however, one exception, namely the distinct discrepancy for the 
orbit with symbolic code `1' (scaled action $s\approx 4.267949$).
The deviation is systematic and appears in the same way if the parameters
of the harmonic inversion procedure (such as signal length etc.) are varied.
This point still needs further clarification.
A possible explanation for the discrepancy may lie in 
the fact that the set of resonances from which the signal was constructed
was not complete, since only resonances near the real axis could be included. 
However, this does not explain why only one orbit is strongly affected.
On the other hand, the error might also be due to the theory of 
Refs.~\cite{Vat94,Vat96,Ros94}, or  its application to the three-disk system.
In fact, the `1' orbit 
is the orbit with the largest contributions from terms of higher order
in $l$ to the sum in (\ref{A1_explizit}).
The contributions from the different $l$ terms
and the converged sum over $l$ of the five shortest orbits
are given in Table \ref{cl1-table}.
For comparison, the last column of Table \ref{cl1-table} shows the
corresponding values following from the amplitudes of
the harmonic inversion results.
The `1' orbit exhibits the largest deviation between the $l=0$ contribution
and the converged sum over $l$, followed by the `0' orbit.
For orbits with a symbol length of 2 or longer, the contributions of higher 
$l$ terms to the amplitude ${\cal A}_{\rm po}^{(1)}$ are already so small 
(due to the large absolute value of the stability eigenvalue $\lambda$) 
that it is impossible to decide whether or not there is a discrepancy
between these terms and the harmonic inversion results.
(Note that this is also true for the period doubling of the orbits `0' 
and `1' in Table \ref{cl1-table}.)
However, the harmonic inversion results for the `0' orbit,
which also shows a relatively large contribution from the $l=1$ term,
are in agreement with the theory. Again, it cannot be explained why only
the `1' orbit is affected (although in this case the reasons might
lie in the special geometrical properties of the `0' orbit).
\begin{table}[p]
\caption{Correction terms $C_l^{(1)}$ (in units of the momentum 
$\hbar k$) and their contributions $C_l^{(1)}/\lambda^l$ to the first-order 
$\hbar$ amplitude (\ref{A1_explizit}) for the five shortest
periodic orbits of the three-disk system with $d=6$.
The values are compared with the results obtained by harmonic inversion 
({\rm hi}).
The orbits are characterized by their symbolic code; their
scaled action $s$ and expanding stability 
eigenvalue $\lambda$ are also given. Note that the maximum
correction to the $l=0$ contribution occurs for the orbit `1', and
is given  by the $l=1$ term.}
\label{cl1-table}
\begin{center}
\begin{tabular}{c||rrr|r||r}
& $l$ & $C_l^{(1)}$ & ${\displaystyle{C_l^{(1)}\over\lambda^l}}$ & 
${\displaystyle\sum_{l=0}^\infty{C_l^{(1)}\over\lambda^l}}$
& ${\displaystyle  
\Biggl[\sum_{l=0}^\infty{C_l^{(1)}\over\lambda^l}\Biggr]_{\rm hi}}$\\
\hline
\hline
       & 0 &  0.625000 &  0.625000 & 0.690360 & 0.6934\\
`0'       
       & 1 &  1.125000 &  0.113648 & \\
$s=4.000000$   
       & 2 & -2.750000 & -0.028064 & \\
$\lambda=9.898979$    
       & 3 &-14.750000 & -0.015206 & \\
\hline
        & 0 &  1.124315 &  1.124315 & 0.843867& 1.055 \\
`1'
        & 1 &  3.661620 & -0.311059 & \\
$s=4.267949$
        & 2 &  4.383308 &  0.031633 & \\
$\lambda=-11.77146$    
        & 3 &  1.162291 & -0.000713 & \\
\hline
        & 0 &       1.250000 &       1.250000  & 1.272357 & 1.259 \\     
$2\times$`0'   
        & 1 &        2.250000 &       0.022962  & \\ 
$s=8.000000$    
        & 2 &      -5.500000 &      -0.000573   & \\   
$\lambda=97.98979$    
        & 3 &     -29.500000 &      -0.000031    &  \\   
\hline
        & 0 &  2.039795 &  2.039795  &  1.989582 & 2.019\\
`01'
        & 1 &  6.278740 & -0.050596 & \\
$s=8.316529$
        & 2 &  5.881196 &  0.000382 & \\
$\lambda=-124.0948$ 
        & 3 & -4.066328 &  0.000002 & \\
\hline
        & 0 &       2.248630 &       2.248630   & 2.301937 & 2.270\\
$2\times$`1' 
        & 1 &       7.323240 &       0.052850    &    \\
$s=8.535898$
        & 2 &       8.766615 &       0.000457    &   \\
$\lambda=138.5672$
        & 3 &       2.324582 &       0.000001   &     \\
\end{tabular}
\end{center}
\end{table}

Concluding our discussion of Figure \ref{1.Ordnung} and Table \ref{cl1-table},
we notice that the harmonic inversion results indeed confirm the validity 
of the $l=0$ approximation to the formula (\ref{g1_explizit2}) for orbits 
with large stability eigenvalues.
On the other hand, the results demonstrate that the theory of higher-order 
$\hbar$ corrections to the Gutzwiller formula still contains
unanswered questions, and further investigations are necessary.

\section{Corrections to the semiclassical eigenvalues}
\label{hbar1}
We now turn to the problem of obtaining corrections to the semiclassical 
eigenvalues of chaotic systems from the $\hbar$ expansion 
(\ref{g_hbar_series}) of the periodic orbit sum.
A general procedure for including higher-order $\hbar$ corrections
in the harmonic inversion scheme was developed in 
Refs.~\cite{Mai99a,Mai98c,Wei00},
where it was applied to the circle billiard as an example of an
integrable system.
In the following, we briefly recapitulate the main ideas 
of the procedure and then
apply the technique to the open three-disk system.

\subsection{Theory}
For periodic orbit quantization, usually only the zeroth-order contributions
${\cal A}_{\rm po}^{(0)}$ to the expanded response function
(\ref{g_hbar_series}), corresponding to the Gutzwiller formula
(or, for integrable systems, the Berry-Tabor formula),
are considered.
In the harmonic inversion scheme for semiclassical quantization
\cite{Mai97b,Mai98,Mai99a}, semiclassical approximations to the 
eigenvalues or resonances are determined by adjusting
the Fourier transform of the principal periodic orbit sum
\begin{equation}
 C_0(s) = \sum_{\rm po}{\cal A}_{\rm po}^{(0)} \delta(s-s_{\rm po})
\end{equation}
to the functional form of the corresponding exact quantum expression 
(i.e., the Fourier transform of the exact response function (\ref{gqm}))
\begin{equation}
\label{Cqmsq}
 C_{\rm qm}(s) = -{\rm i} \sum_j m_j\ {\rm e}^{-{\rm i}k_js} \; ,
\end{equation}
with $k_j$ the eigenvalues or resonances and $m_j$ their multiplicities.

Since for $n\ge 1$ the asymptotic expansion (\ref{g_hbar_series}) of the 
semiclassical response function suffers from the singularities 
at $k=0$, higher-order $\hbar$ terms cannot directly be included in the
harmonic inversion scheme.
Instead, the correction terms to the semiclassical 
eigenvalues can be calculated separately, order by order.
We assume that the $(n-1)^{\rm st}$-order $\hbar$ approximations $k_{j,n-1}$ 
to the exact eigenvalues have already been obtained and the 
$n^{\rm th}$-order approximations $k_{j,n}$ are to be calculated.
In terms of these approximations to the eigenvalues,
the difference between the two subsequent approximations to the quantum
mechanical response function reads
\begin{eqnarray}
     g_{n}(k)
 &=& \sum_j \left({m_j\over k-k_{j,n}+{\rm i}0}
          - {m_j\over k-k_{j,n-1}+{\rm i}0}\right)
     \nonumber \\
 &\approx& \sum_j{m_j\Delta k_{j,n}\over (k-\bar k_{j,n}+{\rm i}0)^2} \; ,
\label{g_n}
\end{eqnarray}
with $\bar k_{j,n}={1\over 2}(k_{j,n}+k_{j,n-1})$ and 
$\Delta k_{j,n}=k_{j,n}-k_{j,n-1}$.
Integration of (\ref{g_n}) and multiplication by $k^n$ yields
\begin{equation}
 {\cal G}_{n}(k) = k^n \int g_{n}(k){\rm d}k
 = \sum_j {-m_jk^n\Delta k_{j,n}\over k-\bar k_{j,n}+{\rm i}0} \; .
\label{g_int_qm}
\end{equation}
The periodic orbit approximation to (\ref{g_int_qm}) is obtained from 
the term $g_{n}(k)$ in the periodic orbit sum (\ref{g_hbar_series}) by 
integration and multiplication by $k^n$, yielding
\begin{equation}
     {\cal G}_{n}(k)
 = -{\rm i}\sum_{\rm po} {1\over s_{\rm po}}
     {\cal A}_{\rm po}^{(n)} {\rm e}^{{\rm i}ks_{\rm po}}
     + {\cal O}\left(1\over k\right) \; .
\label{g_int_sc}
\end{equation}
One can now Fourier transform both (\ref{g_int_qm}) and (\ref{g_int_sc}),
and obtains ($n\ge 1$)
\begin{eqnarray}
     C_{n}(s)
 &\equiv& {1\over 2\pi}\int_{-\infty}^{+\infty}{\cal G}_{n}(k)
     {\rm e}^{-{\rm i}ks}{\rm d}k \nonumber \\
\label{Cn_qm}
 &=& {\rm i}\sum_j m_j (\bar k_j)^n\Delta k_{j,n}{\rm e}^{-{\rm i}\bar k_js} \\
\label{Cn_sc}
 &\stackrel{\rm h.i.}{=}&
     -{\rm i}\sum_{\rm po}{1\over s_{\rm po}}{\cal A}_{\rm po}^{(n)}
     \delta(s-s_{\rm po}) \; .
\end{eqnarray}
Equations (\ref{Cn_qm}) and (\ref{Cn_sc}) imply that the $\hbar$ expansion 
of the semiclassical eigenvalues can be obtained, order by order, by 
adjusting the periodic orbit signal (\ref{Cn_sc}) to 
the functional form of (\ref{Cn_qm}) by harmonic inversion (h.i.).
The frequencies $\bar k_j$ of the periodic orbit signal (\ref{Cn_sc}) are the 
semiclassical eigenvalues or resonances, averaged over different 
orders of $\hbar$.
Note that the accuracy of these values does not necessarily
increase with increasing order $n$.
The corrections $\Delta k_{j,n}$ to the eigenvalues are not obtained
from the frequencies, but from 
the {amplitudes}, $m_j(\bar k_j)^n\Delta k_{j,n}$, of the periodic 
orbit signal.

\subsection{Application to the three-disk scattering system}
We have applied the above procedure to the open three-disk scatterer
with disk separations $d=6$ and $d=2.5$.
For both cases, we first calculated the zeroth-order
$\hbar$ approximations to the resonances and then determined the 
first-order $\hbar$ corrections to the semiclassical results
following the scheme outlined above.

\begin{figure}
\vspace{18.9cm}
\includegraphics{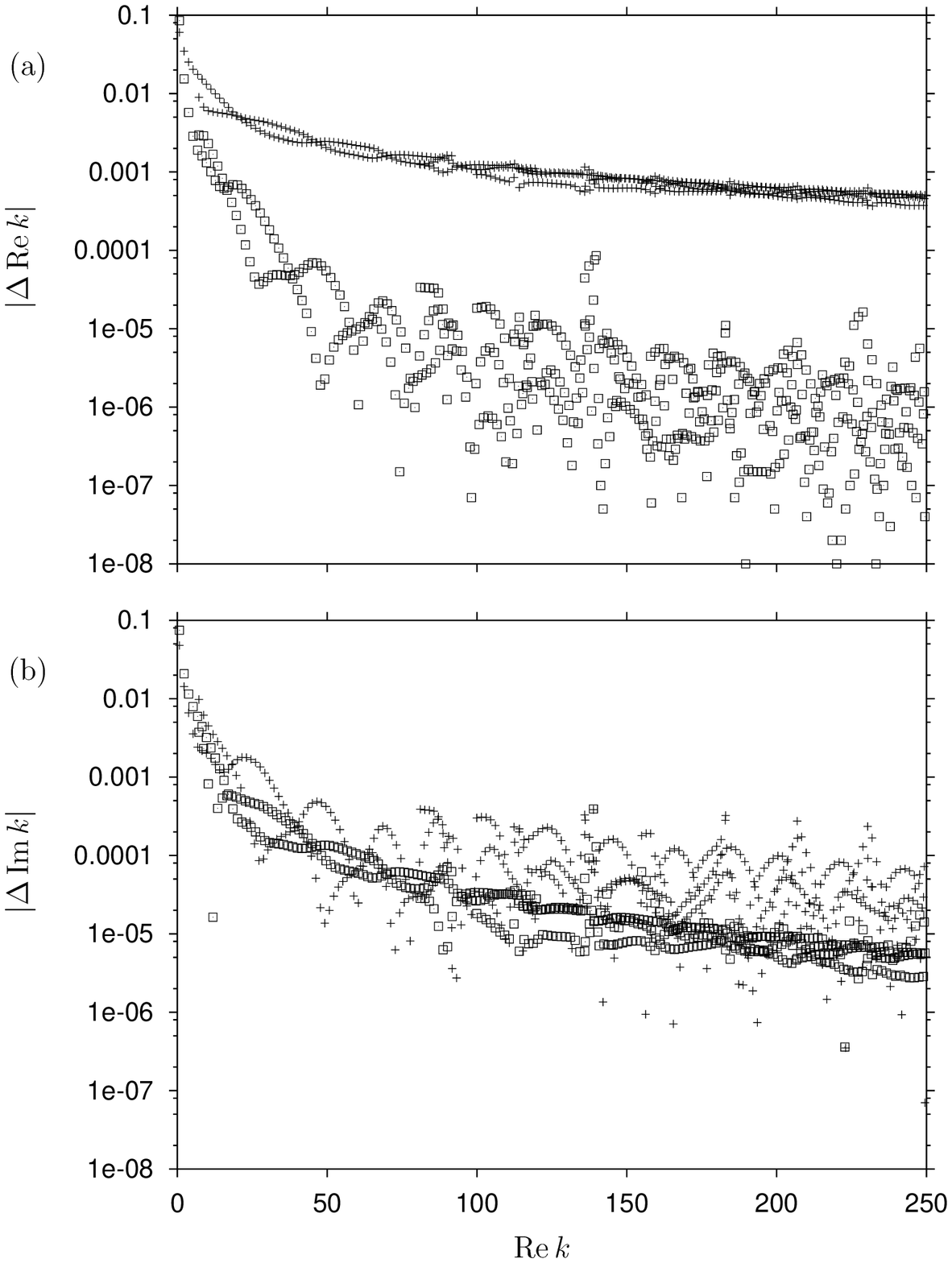}
\caption{The semiclassical errors of the zeroth ($+$) and first 
($\boxdot$) -order approximations to the complex eigenvalues of the 
resonances of the three-disk system with $d=6$, 
plotted as a function of the real part of the resonances. 
Only resonances with  imaginary parts ${\rm Im}\,k\ge -0.5$ are included.}
\label{hbarl0.Re.d6}
\end{figure}

\begin{table}[p]
\caption{Zeroth ($k_0$) and first ($k_1$) -order approximations to 
the complex eigenvalues of the resonances
 of the three-disk system with disk separation $d=6$ ($A_1$ subspace), obtained 
by harmonic inversion of a signal of length $s_{\rm max}=56$. For comparison, 
the exact 
quantum values $k_{\rm ex}$ are given (taken from Refs.~\cite{Wir99,Cvi97,Wir}).
Only resonances of the four leading bands with ${\rm Im}\,k\ge -0.5$ are 
included.}
\label{tab_k1_d6}
\begin{center}
\begin{tabular}{|rr|rr|rr|}
\hline
${\rm Re}\,k_{0}$ & ${\rm Im}\,k_{0}$ & 
${\rm Re}\,k_{1}$ & ${\rm Im}\,k_{1}$ & 
${\rm Re}\,k_{\rm ex}$ & ${\rm Im}\,k_{\rm ex}$ \\
\hline\hline
0.75831  &  -0.12282  &    0.61295  &  -0.14993  &    0.69800  &  -0.07501 \\
2.27428  &  -0.13306  &    2.22417  &  -0.13960  &    2.23960  &  -0.11877 \\
3.78788  &  -0.15413  &    3.75695  &  -0.15903  &    3.76269  &  -0.14755 \\
5.29607  &  -0.18679  &    5.27282  &  -0.19113  &    5.27567  &  -0.18322 \\
6.79364  &  -0.22992  &    6.77417  &  -0.23345  &    6.77607  &  -0.22751 \\
7.22422  &  -0.49541  &    7.21231  &  -0.48189  &    7.21527  &  -0.48562 \\
8.27639  &  -0.27708  &    8.25953  &  -0.27932  &    8.26114  &  -0.27491 \\
8.77919  &  -0.43027  &    8.76958  &  -0.42179  &    8.77247  &  -0.42410 \\
9.74763  &  -0.32082  &    9.73320  &  -0.32201  &    9.73451  &  -0.31881 \\
10.34423  &  -0.37820  &   10.33588  &  -0.37289  &   10.33819  &  -0.37371 \\
\multicolumn{1}{|c}{\vdots} &
\multicolumn{1}{c|}{\vdots} &
\multicolumn{1}{c}{\vdots} &
\multicolumn{1}{c|}{\vdots} &
\multicolumn{1}{c}{\vdots} &
\multicolumn{1}{c|}{\vdots} \\

150.09512  &  -0.23623  &  150.09449  &  -0.23613  &  150.09450  &  -0.23613 \\
150.76086  &  -0.40911  &  150.76004  &  -0.40908  &  150.76004  &  -0.40906 \\
151.09908  &  -0.22292  &  151.09826  &  -0.22298  &  151.09826  &  -0.22297 \\
151.64342  &  -0.22327  &  151.64279  &  -0.22321  &  151.64279  &  -0.22320 \\
152.24814  &  -0.38924  &  152.24733  &  -0.38920  &  152.24733  &  -0.38919 \\
152.60380  &  -0.24729  &  152.60298  &  -0.24735  &  152.60298  &  -0.24733 \\
153.19200  &  -0.21587  &  153.19138  &  -0.21583  &  153.19138  &  -0.21582 \\
153.73475  &  -0.36935  &  153.73395  &  -0.36932  &  153.73395  &  -0.36931 \\
154.11072  &  -0.27186  &  154.10992  &  -0.27192  &  154.10992  &  -0.27190 \\
154.74201  &  -0.21392  &  154.74140  &  -0.21390  &  154.74140  &  -0.21389 \\
\hline
\end{tabular}
\end{center}
\end{table}
For disk separation $d=6$, we used the periodic orbits up to length
$s_{\rm max}=56$ to calculate the resonances in the region 
$0\le {\rm Re}\, k\le 250$.
In the first-order amplitudes (\ref{A1_explizit}), only the leading-order
term $l=0$ of the sum was included. 
(The terms with $l \ge 1$ contribute significantly only to the `1' orbit
and thus basically do not effect the semiclassical resonances.)
The results for the first-order corrections $\Delta k_1$ were added to the 
semiclassical results to obtain the first-order approximations $k_1$ to 
the resonances.
Table \ref{tab_k1_d6} shows part of the results in the regions
${\rm Re}\,k\in[0,12]$ and ${\rm Re}\,k\in[150,155]$ with 
${\rm Im}\,k\ge -0.5$. 
For comparison, the exact quantum
resonances $k_{\rm ex}$ from Refs.~\cite{Wir99,Cvi97,Wir} are also given.
We note that the semiclassical resonances $k_0$ obtained by harmonic
inversion agree with the semiclassical cycle expansion values from
Refs.~\cite{Wir99,Cvi97,Wir} to all digits given.
Figure \ref{hbarl0.Re.d6} compares the 
semiclassical errors 
of the zeroth-order (crosses) and first-order (squares) approximations
as a function of 
the real part of the resonances.
The deviations of the real and imaginary parts from the exact quantum values
are shown separately.
Only resonances with an imaginary
part ${\rm Im}\, k\ge -0.5$ were included in the plot.

The results presented in Figure \ref{hbarl0.Re.d6} show that 
by including the first-order corrections
a significant improvement in the accuracy of the real
parts of the resonances is achieved. This is evident from
Figure \ref{hbarl0.Re.d6}  in spite of the
fact that a one-to-one correspondence between the zeroth and first-order 
values plotted is difficult to establish with the naked eye.
For most resonances, the real part of the first-order approximation lies 
between two and five orders of magnitude closer to the exact quantum
values than the zeroth-order approximation.
Only for the ``most quantum'' resonances, with very low real parts, the 
improvement is rather small. 
The reason for this lies in the nature of the
semiclassical approximation as an approximation itself: 
in order to improve these values, second or higher-order terms of the
$\hbar$ expansion must be considered.

The accuracy of the imaginary parts of the semiclassical resonances
is less significantly increased by the first-order corrections
than that of the real parts.
For some resonances, the zeroth-order approximation lies even closer
to the exact quantum values than the first-order approximation.
This was also observed in Refs.\ \cite{Alo93,Ros94}, where the first-order 
$\hbar$ corrections to the resonances was calculated using the cycle expansion
technique.
As discussed in \cite{Alo93,Ros94}, the first-order corrections to the
periodic orbit sum mainly improve the real part of the resonances, while 
the imaginary part can be expected to be improved by second-order $\hbar$
corrections.

A similar behaviour can be found for disk separation $d=2.5$.
Here, we calculated the semiclassical resonances and
their first-order $\hbar$ corrections in the range
$0\le {\rm Re}\, k\le 90$ and $-0.82\le {\rm Im}\, k\le 0$
from the periodic orbits up to length $s_{\rm max}=12$.
In the first-order amplitudes (\ref{A1_explizit}), again only the 
$l=0$ term was included. Table \ref{tab_k1_d2.5} compares part of the
results for the first-order approximations to the resonances
with the zeroth-order approximations and the exact quantum values.
The zeroth order resonances $k_0$ obtained by harmonic inversion
agree with the cycle expansion
values calculated by Wirzba \cite{Wir99,Cvi97,Wir} (not shown) to at least
four significant digits.
\begin{table}[p]
\caption{Zeroth and first-order approximations to the 
complex eigenvalues of the resonances
of the three-disk system with disk separation $d=2.5$ ($A_1$ subspace),
obtained from a signal of length $s_{\rm max}=12$.
The notations are the same as in Table \ref{tab_k1_d6}.
The table contains the resonances in the region $1\le {\rm Re}\, k\le 90$ and $-0.82\le {\rm Im}\, k\le 0$.}
\label{tab_k1_d2.5}
\begin{center}
\begin{tabular}{|rr|rr|rr|}
\hline
${\rm Re}\,k_{0}$ & ${\rm Im}\,k_{0}$ & 
${\rm Re}\,k_{1}$ & ${\rm Im}\,k_{1}$ & 
${\rm Re}\,k_{\rm ex}$ & ${\rm Im}\,k_{\rm ex}$ \\
\hline\hline
4.58118  &  -0.08999  &   4.35123  &  -0.05580  &   4.46928  &  -0.00157 \\
7.14428  &  -0.81079  &   6.90301  &  -0.66547  &   7.09171  &  -0.72079 \\
13.00009  &  -0.65163  &  12.93645  &  -0.63795  &  12.95032  &  -0.62824 \\
17.57004  &  -0.68486  &  17.45278  &  -0.65154  &  17.50423  &  -0.63526 \\
18.92585  &  -0.78389  &  18.93139  &  -0.72879  &  18.92545  &  -0.76629 \\
27.88820  &  -0.54319  &  27.86253  &  -0.55690  &  27.85779  &  -0.54993 \\
30.38846  &  -0.11345  &  30.34790  &  -0.11469  &  30.35289  &  -0.10567 \\
32.09670  &  -0.62237  &  32.05975  &  -0.61112  &  32.06937  &  -0.60774 \\
36.50664  &  -0.38464  &  36.48222  &  -0.38774  &  36.48228  &  -0.38392 \\
39.81392  &  -0.35801  &  39.78247  &  -0.35590  &  39.78597  &  -0.35087 \\

\multicolumn{1}{|c}{\vdots} &
\multicolumn{1}{c|}{\vdots} &
\multicolumn{1}{c}{\vdots} &
\multicolumn{1}{c|}{\vdots} &
\multicolumn{1}{c}{\vdots} &
\multicolumn{1}{c|}{\vdots} \\

65.68047  &  -0.27378  &  65.66353  &  -0.27480  &  65.66387  &  -0.27258 \\
67.86889  &  -0.28815  &  67.85047  &  -0.28896  &  67.85151  &  -0.28656 \\
69.34446  &  -0.31247  &  69.33251  &  -0.30929  &  69.33346  &  -0.30925 \\
71.08294  &  -0.53828  &  71.06684  &  -0.53676  &  71.06727  &  -0.53534 \\
74.85524  &  -0.30224  &  74.83975  &  -0.30093  &  74.84053  &  -0.29941 \\
77.31939  &  -0.31303  &  77.30827  &  -0.31116  &  77.30881  &  -0.31071 \\
80.41789  &  -0.36657  &  80.39883  &  -0.36525  &  80.40022  &  -0.36289 \\
81.69995  &  -0.56162  &  81.68874  &  -0.55515  &  81.69091  &  -0.55547 \\
83.87557  &  -0.50399  &  83.86231  &  -0.50159  &  83.86311  &  -0.50054 \\
85.80058  &  -0.41490  &  85.79208  &  -0.41566  &  85.79189  &  -0.41529 \\

\hline
\end{tabular}
\end{center}
\end{table}

\begin{figure}
\vspace{19.2cm}
\includegraphics{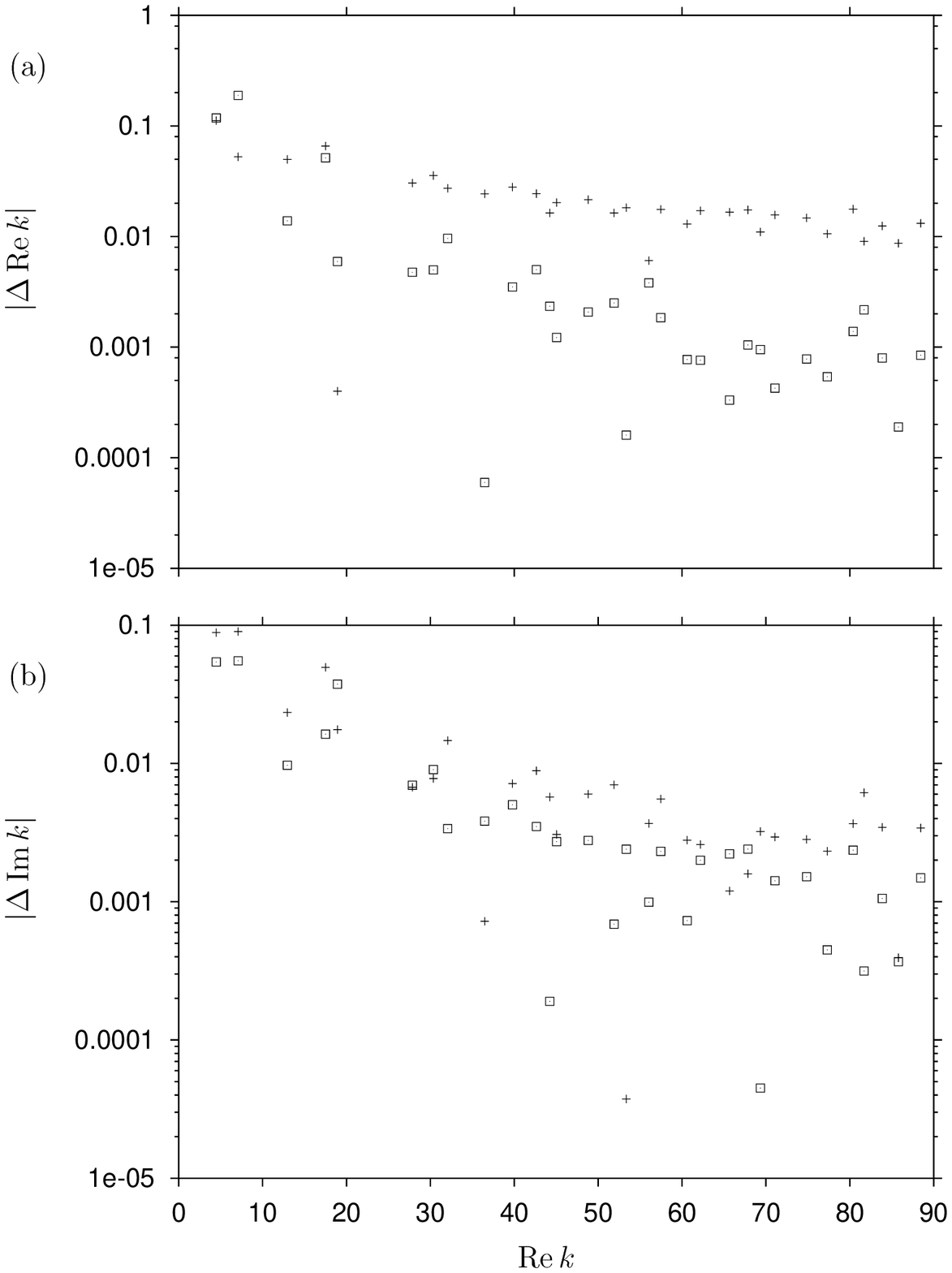}
\caption{As Fig.~\ref{hbarl0.Re.d6}, but for disk separation $d=2.5$.
Only resonances with  imaginary parts ${\rm Im}\,k\ge -0.82$ are included.}
\label{hbarl0.Re.d2.5}
\end{figure}

Again, we determined the semiclassical error of the first-order
approximations to the resonances in comparison to that of the zeroth-order 
approximation.
The results are presented in Figure \ref{hbarl0.Re.d2.5}. 
The general behaviour of the values is similar to that in the case $d=6$ 
discussed above, although the improvement of the accuracy achieved by the 
first-order corrections is not as spectacular as for $d=6$. 
The reason for this may partly lie in the error induced by the harmonic 
inversion method, which for $d=2.5$ is larger already in the zeroth-order 
approximation than for $d=6$. 
The results could in principle be improved by extending the signal to 
longer orbits.
On the other hand, in the part of the spectrum considered, second and 
higher-order $\hbar$ corrections may be more important than in the case $d=6$.
However, it is evident from Figure \ref{hbarl0.Re.d2.5} that,
apart from the resonances with very small real parts, the semiclassical error 
of the real parts of the resonances could still be reduced, even for the small
disk separation of $d=2.5$, by the first-order $\hbar$ corrections by one or 
two orders of magnitude.

\section{Conclusions}
In this paper we have demonstrated the power of the harmonic inversion
technique in explicitly determining higher-order $\hbar$ corrections to the 
Gutzwiller trace formula and to the semiclassical eigenvalues
of a completely chaotic system, namely the three-disk scattering system.
The method has been used in two directions: 
(1) for the harmonic analysis of the exact quantum spectrum, 
(2) for the direct calculation of  higher-order corrections to the
semiclassical eigenvalues from classical periodic orbit data.

The harmonic analysis of the exact quantum spectrum of the three-disk system 
with the ``standard'' literature disk separation of $d=6$ first yielded the 
zeroth-order semiclassical amplitudes of the periodic orbit sum
(i.e., the amplitudes entering the Gutzwiller formula), which 
were found to be in perfect agreement with the Gutzwiller amplitudes
calculated directly from classical periodic orbit data. 
Next, from the exact quantum resonances
and their zeroth order approximations, we were able to compute the first-order 
amplitudes applying harmonic inversion to Eq.~(\ref{G_n}).
We could verify the correctness of the values  obtained in this way by 
comparing with the results of an alternative theoretical approach  
\cite{Vat94,Vat96,Ros94} for calculating first-order corrections to the 
Gutzwiller formula in chaotic billiards, which we implemented for 
the three-disk system. The results turned out to be in very good agreement 
(on the order of 1.5 per cent, or better), with one notable exception, 
namely the `1' orbit, for which a distinct discrepancy 
(on the order of 20 per cent) persisted. We have discussed
possible origins of the discrepancy although an ultimate reason could not
be identified. Therefore, in spite of the 
very good agreement of the results  in all other cases, we have to conclude 
that the theory of $\hbar$ corrections
to the Gutzwiller formula still contains unanswered questions. [We note
that in fact for {\em integrable} systems there does not yet exist
a general theory for higher-order $\hbar$ corrections to the Berry-Tabor
formula at all.]

In the direct calculation of  higher-order corrections to semiclassical
eigenvalues from classical periodic orbit data, we first 
evaluated  the 
first-order correction amplitudes to the Gutzwiller formula
as given by the theory of Vattay and Rosenqvist, and then,
by harmonic inversion  of Eq.~(\ref{Cn_sc}),  determined the first-order
$\hbar$ corrections to the semiclassical (complex) eigenvalues of
resonances of the three-disk scattering system with disk separations
$d=6$ and $d=2.5$. 
For both distances, the semiclassical error, as compared to the exact quantum
values, of the zeroth-order results (obtained from the Gutzwiller formula
by harmonic inversion)
for the real parts of the resonances
could be significantly reduced by including the first-order 
$\hbar$ corrections: the accuracy was
increased by two to five orders of magnitude for $d=6$, 
and still by one to two orders of magnitude for $d=2.5$. 
Only for the ``most quantum'' resonances, with very small real parts,
the increase in accuracy was found to be rather modest. 
It turned out that the accuracy of the imaginary parts of the semiclassical 
eigenvalues of resonances was less significantly increased
by  the first-order corrections; here, second-order corrections
would have to be considered.

Although in our calculations we have used literature values for the 
semiclassical resonances and for the zeroth and first-order amplitudes, 
we could have performed, in principle, all calculations knowing the exact 
quantum resonances only.
The analysis of the exact quantum spectrum yields the semiclassical
amplitudes, which in turn can be used to calculate the semiclassical
resonances. Then, by an analysis of the difference spectrum between 
semiclassical and exact resonances, the first order amplitudes can be 
determined, which again can be used to obtain the first-order corrections
to the resonances.
Although we have concentrated in our examples on obtaining  first-order
$\hbar$ corrections, it is evident from our discussion that the next-order 
corrections could be obtained iteratively in an analogous manner
by repeated application of Eqs.~(\ref{G_n}), and (\ref{Cn_qm}), 
(\ref{Cn_sc}), respectively. 

In summary, we have demonstrated that harmonic inversion
-- as a means for circumventing the convergence
problems of semiclassical trace formulae -- 
is indeed a very efficient and universal tool, not only for semiclassical
quantization, but also for the explicit calculation of higher-order $\hbar$ 
corrections to the semiclassical eigenvalues or resonances even 
in chaotic systems. Moreover, harmonic inversion does not rely
on specific assumptions for the systems under consideration, and therefore
an application of the methods presented in this paper to other chaotic 
systems will be promising and worthwhile.

\begin{acknowledgement}
We are grateful to G.~Vattay for sending us the code for the calculation of
the first-order corrections to the Gutzwiller formula for chaotic billiards.
We thank A.~Wirzba for communicating to us quantum mechanical
and semiclassical resonances of the three-disk system. This work was
supported by Deutsche Forschungsgemeinschaft and Deutscher Akademischer
Austauschdienst.
\end{acknowledgement}


\end{document}